\documentclass[apj]{emulateapj}
\usepackage{url}
\usepackage{amsmath}
\slugcomment{Accepted to PASP}
\begin{document}

\title{Interstellar Dust models towards some IUE stars}
\author{N. Katyal$^{1,2}$, R Gupta$^{1}$, D B Vaidya$^{3}$}
\affil{$^{1}$Inter University centre for Astronomy and Astrophysics, Pune, India\\
$^{2}$School of Physical Sciences, Jawaharlal Nehru University, New Delhi 110067, India \\
             $^{3}$ICCSIR, Ahmedabad, 380006, India}
\received{August 16, 2003}\revised{September 17, 2003}\accepted{September 30, 2003}
\begin{abstract}

We study the extinction properties of the composite dust grains, consisting of host silicate spheroids and graphite as inclusions, using discrete dipole approximation (DDA). We calculate the extinction cross sections of the composite grains in the ultraviolet spectral region, 1200\AA~-3200\AA~and study the variation in extinction as a function of the volume fraction of the inclusions. We compare the model extinction curves with the observed interstellar extinction curves obtained from the data given by the International Ultraviolet Explorer (IUE) satellite. Our results for the composite grains show a distinct variation in the extinction efficiencies with the variation in the volume fraction of the inclusions. In particular, it is found that the wavelength of peak absorption at `2175\AA' shifts towards the longer wavelength with the variation in the volume fraction of inclusions. We find that the composite grain models with the axial ratios viz. 1.33 and 2.0 fit the observed extinction reasonably well with a grain size distribution, a =  0.005-0.250$\mu m$.  Moreover, our results of the composite grains clearly indicate that the inhomogeneity in the grain structure, composition and the surrounding media modifies the extinction properties of the grains.
\end{abstract}
\keywords{Interstellar Dust, Extinction, Ultraviolet spectra}

\section{INTRODUCTION}
In 1968, the first satellite, OAO2, capable of UV observations was launched. Till then, due to atmospheric extinction, the astronomical studies in the ultraviolet spectral region were not possible. Later on other satellites like TD-1, Astronomical Netherland satellite (ANS) and International Ultraviolet Explorer (IUE) were launched. IUE has provided a wealth of data on interstellar extinction in the UV region. We study the wavelength dependence of the interstellar extinction in the UV region, 1200-3200\AA~, observed by the IUE satellite towards several stars in the various interstellar environments; viz. diffuse interstellar medium, HII region, OB type association, reflection nebulae, dense medium and HI sources. 

Recent studies of interplanetary, cometary and interstellar dust indicate that the cosmic dust grains are inhomogeneous viz. porous, fluffy and composite. The collected interplanetary particles are also porous and composite \citep{brown1987}. 
\citet{mathis96}, \citet{dwek97}, \citet{greenli98} and \citet{zubko94} 
have proposed composite grain models consisting of silicates and amorphous carbon to explain the observed wavelength dependence of interstellar extinction, polarization, albedo, IR emission and the observed
elemental depletion. They have used the effective medium theory (EMT). \citet{iati04} have
studied optical properties of the composite grains using the transition matrix
approach. \citet{vosh06} and \citet{vosh08} have
used the layered sphere method to study the extinction properties of the porous
grains. Very recently, \citet{sieb2013} have used a dust model, consisting of a mixture of large spheroidal amorphous carbon (AMC) and silicate grains. Small grains of graphite, silicates and polycyclic aromatic hydrocarbons (PAHs) are also included to explain the extinction, emission, linear and circular polarization in the diffuse interstellar medium. \citet{clayton03MEM} have used Maximum Entropy Method (MEM) and EMT for  
2-component (silicates and graphite) and 3-component (silicates, graphite and 
amorphous carbon) spherical grain models to study the extinction properties in the 
Milky Way galaxy and the Magellanic clouds. In EMT, the inhomogeneous particle is replaced
by a homogeneous one with some `average effective dielectric function'. The
effects related to the fluctuations of the dielectric function within the inhomogeneous
structure can not be treated by this approach of the EMT.

We have used Discrete Dipole Approximation (DDA) which allows consideration for irregular shape effects,
surface roughness and internal structure within the grain \citep{wolff94,wolff98}. Since there is no exact theory to study these porous and composite particles, there is a need to formulate models of electromagnetic
scattering by approximate methods like EMT and DDA. We have used DDA to calculate the extinction cross sections of the composite grains in the spectral UV region, 1200\AA~ - 3200\AA~,  and compared the model extinction curves with the extinction curves, derived from the IUE satellite
observations. For a discussion and comparison of EMT and DDA see for e.g. \citet{Ossenkopf91} and \citet{wolff98}. Earlier \citet{vaidya01,gupta07} have used composite grain models to interpret the average observed interstellar extinction. Moreover, the recent results of \citet{katyal11} show that the composite grain model is more efficient as compared to the bare silicate/graphite grain models in producing the extinction and also reducing the cosmic abundance constraints. Composite dust grain models are also being employed to analyze IR emission. Recently, \citet{vaidya11} have used the composite grain model to interpret the observed IR emission from circumstellar dust.  
\citet{massa83} have done spectrophotometric measurements for a sample of stars judged 
by \citet{meyer81} to study highly anomalous peculiar UV extinction as inferred from 
the broad-band Astronomical Netherlands Satellite (ANS) data. These observations showed a discrete bump feature at 2175\AA~ \citep{stecher65,stecher69}.
This feature has been ascribed to small graphite 
grains \citep{stecherdonn65,draine93}.
 
Other possible candidate for the spectral bump at 2175\AA~ could be polycyclic aromatic hydrocarbons (PAHs) as discussed by \citet{li2001} and \citet{malloci2008}. \citet{sieb2013} have also discussed about the strong electronic transitions of both graphites and PAHs at 2175\AA~ to be responsible for the bump feature. \citet{green83} have 
found strong correlation between the strength of the `2175\AA~' feature
and the visible extinction. They obtained a poor correlation between far ultraviolet 
(FUV) extinction, strength of the feature and visible extinction concluding that 
a wide spectrum of size distribution is needed to explain the average observed 
interstellar extinction curve. \citet{xiang2011} have shown that the carriers responsible for the 2175\AA~ feature and the extinction in the UV might not be the same.  

Wavelength dependent studies of the interstellar extinction curves are the best tool for understanding environment around these stars. The most commonly used technique for deriving the wavelength dependence
of interstellar extinction is the ``pair method" \citep{massa83}.
Basically, the ratio of the fluxes of the reddened and comparison star gives a direct measurement of the dust extinction towards the reddened star. The resultant ratio, after
normalization is referred to as the `extinction curve'. Errors resulting from poorly
matched pairs can dominate the uncertainties of individual extinction curves.
\citet{massa86,massa88,fitz90} analyzed several IUE extinction curves
and found that all these curves could be fitted extremely well by a single
analytical expression with six parameters. 

With the availability of much more observational data, revisions of earlier dust models was done as extinction of light is highly subject to the interstellar environments from where it passes through. Therefore, \citet{clayton88,clayton89} (hereafter called CCM method) found that in general, the properties of UV extinction curves are correlated with the extinction in the optical/IR region and that from the UV through the IR. They characterized this dependence by a single parameter, $R_{v}^{-1}$, which is the ratio of visual extinction to total extinction of V, and is defined as $R_{v}$=A(V)/E(B-V).
However, the CCM method has its limitations,
both from the standpoint of understanding dust grain properties and dereddening
energy distributions. While the UV curve shapes indeed correlate in general
with $R_{v}$, the $R_{v}^{-1}$ dependence adopted by CCM is insufficient to describe the
behavior over the entire range of observed $R_{v}$ values, and breaks down at small $R_{v}$.
Further, the CCM formula does not provide particularly good fits to individual extinction
curves. Evidently, factors other than $R_{v}$, e.g. chemical composition, processing
history, ambient radiant field play important roles in determining the extinction
properties. Hence, based on different interstellar environments of the stars, \citet{aiello} have presented a collection of 115 extinction curves derived from low dispersion IUE spectra. 
The atlas includes extinction originating in the diffuse medium and several major 
nebulae and dense clouds. The data can be easily accessed and used for various extinction studies.

The shape of extinction curves are substantially different for different $R_{v}'s$, 
and hence changes in the size distributions is also expected. 
As \citet{cardelli91} have pointed out, lines of sight with large $R_{v}$ are ideal 
for examining processes that modify the grain properties in dense clouds. A good correlation between the strength of the `2175\AA' UV bump feature and the visual extinction was also noted by \citet{green83}. 
\citet{Wein} have constructed size distributions for spherical carbonaceous and silicate grain 
populations in different regions of the Milky way, LMC and SMC to account for the observed near IR and microwave emission from diffuse interstellar environment using a fairly simple functional form, characterized by various parameters. They have shown that these variations can be well parameterized by $R_{v}$. 
Another study by \citet{kim94} found out that denser environments with high $R_{v}$(=5.3) have the presence of larger mean size of grains, though all denser regions may not necessarily have high $R_{v}$. 

In the present study, we have used the `Pair method' which is described in the section 2.1. The main purpose of the present study is to infer the size distribution, shape and composition of the interstellar dust grains, in various interstellar environments (for different values of $R_{v}$), which are consistent with the observed extinction. We use composite grain models to compare extinction toward these stars as observed by the IUE satellite. We tabulate a list of the selected stars and describe the pair method to generate the extinction curves in the UV spectral regime for these stars in section 2. DDA technique and the generation of composite grain models using it are illustrated in Section 3. In section 4, we give the results of the computed model curves and compare these model extinction curves with the observed extinction curves. In section 4, we analyze these results in detail and compare our results in terms of size and composition with those obtained by others. Our conclusions from this study are summarized in section 5.

\section{Preliminary data reduction}

\subsection{PAIR METHOD}

The standard Pair method technique is used for a set of IUE stars to generate the extinction curves. 
The technique involves selecting a highly reddened star 
and comparing it with a star (flux standard) which has negligible reddening and
whose spectral features closely match with those of the reddened star. 
An extinction curve is then constructed by the standard relation (\citet{massa83}):

\begin{equation}
\frac{E(\lambda-V)}{E(B-V)}=\frac{m(\lambda-V)-m(\lambda-V)_{o}}{(B-V)-(B-V)_{o}}
\end{equation}

where subscript `$o$' refers to the unreddened star and other is for the reddened star. 
Here E(B-V) is the difference in extinction between the specified wavelengths and 
corresponds to the color excess. The resultant extinction curve $E(\lambda-V)/E(B-V)$
is then plotted versus $1/\lambda$ for selected IUE stars.

\subsection{Object Selection Criteria}
 
\noindent
We have selected 26 ``program stars" (listed in Table 1)
from \citet{massa88}, \citet{fitz09} and IUE spectral atlas by \citet{wu}. The $R_{v}$ values of the sample reddened stars are taken from \citet{valencic04}. 
The spectral types for these 26 stars lies in the range O7-B5. We have selected 
reddened and dereddened stars on the basis of their visible spectral type and the 
luminosity class. Spectral type mismatch error larger than one 
luminosity subclass is avoided (see Table 1) in order to account for spectral type 
uncertainties between reddened and the dereddened stars. Late type stars are excluded because their ultraviolet energy distributions are very strong functions of their spectral type - thus amplifying the magnitudes of error associated with the spectral mismatching between the reddened and unreddened stars. 
\citet{massa83} and \citet{massa84} give the identification of the 
features useful in matching B stars near the main sequence. Most of the sample stars are selected along different line of sights and are previously known to produce extinction curves that vary considerably from the average Milky way curve ($R_{v}$=3.1). 
The lowest value of 
color excess E(B-V) for unreddened stars sample is 0.01 and the highest value of E(B-V) 
for reddened stars sample is 0.95. The stars selected represent a range of environments;
viz. diffuse interstellar medium (DIF); HII region (HII); OB type association (OB);
reflection nebulae (RN); dense medium (DEN) and radio or HI source (H/RADIO) which are mentioned in second column of Table 1. Environment type for the stars is taken from \citet{massa88,green93} and SIMBAD astronomical database. It is to be noted that the sample of stars selected span the
value of R$_{v}$ i.e the ratio of total to selective extinction, from $\sim$ 2.0 to 5.0 (see
Table 1) representing the physical environments in the galaxy i.e 
from a diffuse medium to a very dense medium so as to study the effects 
on the corresponding extinction curves.  Other properties of the sample stars such as distance (in Kpc) and neutral hydrogen column densities for the sample stars chosen are given in \citet{fitz90}. 

The column (1) of Table 1 gives the HD number of the Program star, column (2) refers to the environment type, column (3) gives 
HD number of comparison star followed by their visible spectral types, column (4) and (5) give 
magnitude in V and B band respectively, column (7) gives the color excess E(B-V) values and column (8) 
gives the value of observed $R_{v}$.

\noindent

\begin{table*}
	\caption{Extinction curve details for the program stars.} 
\begin{tabular}{l l l c c c c  c }
\hline
HD \# (Sp Type) & ENV$^a$ &Flux Std (Sp Type)  & V & B & B-V & E(B-V) & $R_{v} $ \\
\hline
239693 (B5 V) &DIF        & 25350  (B5 V)   & 9.54 & 9.77 & 0.23 & 0.41 & 2.37   \\
185418 (B0.5V)&DEN       & 55857  (B0.5 V) & 7.45 & 7.67 & 0.22 & 0.50 & 2.54    \\
123335 (B5 IV)&HII        & 147394 (B5 IV)  & 6.31 & 6.37 & 0.06 & 0.24 & 2.60     \\
18352  (B1 V) &DIF        & 31726  (B1 V)   & 6.80 & 7.03 & 0.23 & 0.47 & 2.66 \\
54439  (B1.5V)&HII        & 74273  (B1.5V)  & 7.72 & 7.72 & 0.00 & 0.29 & 2.73\\
179406 (B3 V) &HII        & 190993 (B3 V)   & 5.33 & 5.46 & 0.13 & 0.35 & 2.73\\
24432  (B3 II)&HII        & 79447  (B3 III) & 6.93 & 7.51 & 0.58 & 0.51 & 2.77\\
217086 (O7 V) &OB         & 47839  (O7 Vf)  & 7.65 & 8.28 & 0.63 & 0.95 & 2.80\\
46660  (B1 V) &HII        & 31726  (B1 V)   & 8.04 & 8.35 & 0.31 & 0.56 & 2.82\\
281159 (B5 V) &HI/Radio   & 25350  (B5 V)   & 8.53 & 9.21 & 0.68 & 0.86 & 2.85\\
21483  (B3 III)&DEN       & 79447  (B3 III) & 7.03 & 7.38 & 0.35 & 0.55 & 2.89\\
53974  (B0.5IV)&RN        & 149881 (B0.5IV) & 5.38 & 5.41 & 0.03 & 0.31 & 2.94\\
38131  (B0.5V)&RN         & 55857  (B0.5 V) & 8.19 & 8.40 & 0.21 & 0.49 & 3.01\\
217061 (B1 V) &RN         & 31726  (B1 V )  & 8.77 & 9.46 & 0.69 & 0.95 & 3.03\\
205794 (B5 V) &RN         & 25350  (B5 V)   & 8.43 & 8.77 & 0.34 & 0.62 & 3.09\\
46202  (O9 V) &DIF        & 38666  (O9.5 V) & 8.20 & 8.36 & 0.16 & 0.48 & 3.12\\
216658 (B0.5V)&RN         & 55857  (B0.5 V) & 8.89 & 9.59 & 0.70 & 0.98 & 3.14\\
149452 (O9 V) &RN         & 214680 (O9 V)   & 9.07 & 9.65 & 0.58 & 0.84 & 3.37\\
34078  (O9.5V)&DEN        & 38666  (O9.5 V) & 5.99 & 6.18 & 0.19 & 0.54 & 3.42\\
37367  (B2.5V)&DIF        & 37129  (B2.5 V) & 5.98 & 6.11 & 0.13 & 0.40 & 3.55 \\
252325 (B1 V) &RN         & 31726  (B1 V)   &10.79 & 11.36& 0.57 & 0.87 & 3.55\\
147701 (B5 V) &DEN        & 4180   (B5 III) & 8.36 & 8.92 & 0.56 & 0.76 & 3.86\\
147889 (B2 IV)&DEN        & 3360   (B2 IV)  & 7.10 & 7.42 & 0.32 & 1.10 & 3.95\\
37903  (B1.5V)&RN         & 74273  (B1.5 V) & 7.84 & 7.91 & 0.07 & 0.35 & 4.11\\
37061  (B1 IV)&Or N       & 34816  (B1 IV)  & 6.83 & 7.09 & 0.26 & 0.56 & 4.29\\
93222  (O7IIIf)&OB        & 47839  (O7 Vf)  & 8.10 & 8.15 & 0.05 & 0.37 & 4.76\\

\hline
\multicolumn{8}{l}{$^a$  DIFF, Diffuse interstellar medium; DEN, Dense interstellar medium;} \\ 
\multicolumn{8}{l}{HII, HII region; OB, OB association; RN, Reflection nebula }\\
\multicolumn{8}{l}{Or N, Orion Nebula; HI/Radio source.}\\
\multicolumn{8}{l}{ENV type are taken from \citet{massa88,green93} and SIMBAD astronomical} \\
\multicolumn{8}{l}{database.}
\end{tabular}
\end{table*}

Table 2 gives the observational data for the flux standards which are the comparison stars.

\begin{table*}
   \caption{Observational data for flux standard stars}
\begin{tabular}{l l c c c c }
\hline
HD \# &  Sp type & V & B & B-V & E(B-V)\\
\hline
47839 & O7 Vf & 4.65 & 4.41 & -0.24 & 0.08\\
214680& O9 V  & 4.88 & 4.64 & -0.24 & 0.11\\                   
38666 & O9.5 V& 5.17 & 4.89 & -0.28 & 0.02\\   
55857 & B0.5 V& 6.11 & 5.85 & -0.26 & 0.02\\
63922 & B0 III& 4.11 & 3.92 & -0.19 & 0.11\\
149881& B0.5 IV&7.00 & 6.84 & -0.16 &0.09\\
75821 & B0 IV & 5.09 & 4.88 & -0.21 &0.08\\
36512 & B0 V  & 4.62 & 4.36 & -0.26 &0.04\\
34816 & B1 IV & 4.29 & 4.02 & -0.27 & 0.01\\
31726 & B1 V  & 6.15 & 5.94 & -0.21 & 0.05\\
74273 & B1.5 V& 5.87 & 5.69 & -0.18 & 0.03\\
3360  & B2 IV & 3.66 & 3.46 & -0.20 & 0.04\\ 
37129 & B2.5 V& 7.13 & 6.99 & -0.14 & 0.07\\
79447 & B3 III& 3.96 & 3.77 & -0.19 & 0.01\\
190993& B3 V  & 5.07 & 4.89 & -0.18 & 0.02\\ 
147394& B5 IV & 3.90 & 3.75 & -0.15 & 0.01\\ 
25350 & B5 V  & 5.28 & 5.13&-0.15&0.01\\

\hline
\end{tabular}
\end{table*}

\subsection{MERGING OF SPECTRAL BANDS for Pair Method}

Each spectra of program star consisted of two separate images, 
one for the shorter wavelength and another for the longer wavelength. 
Data was taken from following cameras: Short Wavelength Prime 
(SWP,1150\AA~ $< \lambda < 1978$\AA~), Long Wavelength Redundant 
(LWR,1851\AA~ $< \lambda < 3348$\AA~) and Long Wavelength Prime 
(LWP,1851\AA~ $< \lambda < 3347$\AA~). The spectra of each reddened and unreddened star is taken from the IUE archives. For each program  and the comparison star, SWP and 
LWR/LWP data for fluxes were merged to achieve the instrumental resolution i.e ~6\AA~ and the resultant
fluxes are converted to magnitudes $m(\lambda)$, with $\lambda$ 
covering the wavelength range 1150\AA~-3348\AA~. 
The magnitudes were further interpolated in the range 1153-3201\AA~ with a 
binning of 1\AA~. Further, extinction curves are generated using standard Pair method as discussed in Section 2.2.

\section{Discrete Dipole Approximation (DDA) }
An approximate technique called discrete dipole approximation (DDA) was proposed by \citet{purcell73}. It is a powerful numerical technique for calculating the optical properties such as absorption and scattering of the target. DDA is basically designed for targets having arbitrary and irregular shape. As an approximation, the continuum target is replaced by an array of $N$ dipoles. To each dipole, a polarizibility can be assigned for a particular target composition. The polarizibility depends in general on the dielectric properties such as complex refractive index $m=n+ik$ of the material inside the target. The polarizibility and the refractive index of the material can be related to each other by the well known Clausius-Mossotti condition. Each dipole interacts with the neighboring dipoles on the application of electric field. After evaluating the polarization $P$ by all the $N$ dipoles inside the target, we can solve for the absorption and extinction cross sections of the target. The two criteria for the validity of DDA:

1) The value of $\mid m \mid kd$  should be $<$ 1, where $m$ is the complex refractive index of the material, k=$2\pi /\lambda$ is the wave number in vacuum and $d$ is the dipole spacing between the dipoles.

2) The dipole spacing $d$ should be small enough so that the number of dipoles $N$ should be large enough to describe the target shape satisfactorily.

For more detailed calculations, see \citet{draine88}. We have used discrete dipole scattering version 6.1 (DDSCAT6.1$\footnote{http://code.google.com/p/ddscat}$) for the present study. The work by \citet{manual7} may be looked upon for a more detailed analysis on the code.

\subsection{Composite grain models using DDA}
For this study, we have used the DDSCAT6.1 code \citep{draineflat03} which
has been modified and developed by \citet{dobbie99} to generate the composite grain
models. The code, first carves out an outer sphere (or spheroid) from a lattice
of dipole sites. Sites outside the sphere are vacuum and sites inside are
assigned to the host material. Once the host grain is formed, the code locates
centers for internal spheres to form inclusions. The inclusions are of a single
radius and their centers are chosen randomly. The code then outputs a three
dimensional matrix specifying the material type at each dipole site which is then
received by the DDSCAT program. In the present study the axial ratios (hereafter called AR) of the composite spheroidal grains is taken to be AR=1.33, 2.0 and 1.44 with number of dipoles N=9640, 14440 and 25896 respectively. The dipole sites are either silicates, graphites or vacuum. The optical constants of silicates and graphites are taken from \citet{draine85} and \citet{draine87}.
The spheroidal composite grain consists of silicates as the host and graphites as the inclusion. In order to study the effect of volume fraction of graphite inclusion, we use three different volume fraction `$f$' of graphite grain inclusion viz. $f$=0.1, 0.2 and 0.3.  

Table 3 shows the number of dipoles for each grain model along 
with the axial ratio and number of dipoles per inclusion with the number of inclusions 
for each fraction. The calculations of extinction cross sections of the target depend in general upon the orientation of the target. Hence, we average over 27 orientations of the target for all the calculations done by DDA. For more details on the composite grain models and the modified code see \citet{gupta07}.

\begin{table}[!ht]
\caption{Number of dipoles for each inclusion of the grain model along with axes lengths for spheroid in x,y,z direction for host (H) and inclusion (I). Also, number of inclusions is mentioned in brackets in column 3,4 \& 5 for each of the volume fraction $f$ of inclusions.}
\small
\begin{tabular}{lllll}
\\
\hline
N (AR)& $N{x}/N_{y}/N_{z}$  & \multicolumn{3}{c}{No. of dipoles per inclusion}   \\
 & &\multicolumn{3}{c}{(No. of inclusions)}\\
& & f=0.1 & f=0.2 & f=0.3\\
\hline
9640(1.33) &H:32/24/24& 152(6) & 152(11) & 152(16) \\
&   I: 8/6/6 & & &     \\
25896(1.50) & H:48/32/32  & 432(7) & 432(13) & 432(19)  \\
&   I:12/8/8 & & &\\
14440(2.00) &H:48/24/24  &224(6) & 224(11) & 224(16) \\
&   I:12/6/6 & & &\\
\hline
\\
\end{tabular}\\
\end{table}
Fig. 1 illustrates a composite grain model with N=9640 dipoles composed of silicates as host (in green) and graphite as inclusion (in red). The inclusions can be seen clearly in Fig. 2. There are eleven such inclusions consisting of 152 dipoles per inclusion. This model represents a composite dust grain with volume fraction of graphite $f=0.2$.

\begin{figure}[!ht]
\centering
\includegraphics[height=7.4cm]{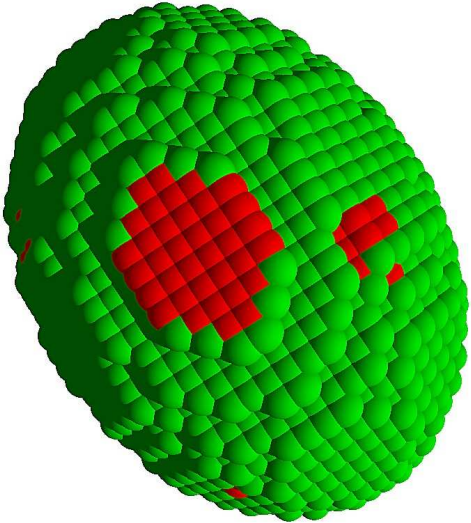}
    \caption{A non-spherical composite dust grain consisting of host (in green) and inclusion (in red) with a
total of N=9640 dipoles where the inclusions embedded in the host spheroid are shown such that only the ones placed at the outer periphery are seen. }
\end{figure}

\begin{figure}[!ht]
\centering
\includegraphics[height=7.4cm]{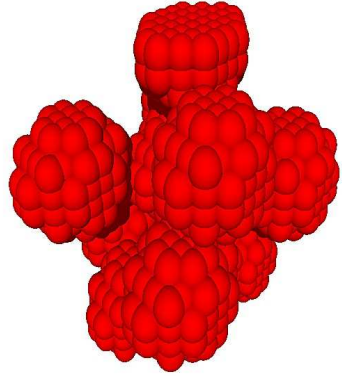}
     \caption{This figure shows the inclusions of the composite grain. The volume fraction $f$ of graphite inclusions is 0.2. The number of inclusions are 11 with 152 dipoles per inclusions.}
\end{figure} 

\section{RESULTS AND DISCUSSIONS}
The following are the principal results of this work:
\subsection{Extinction efficiencies of the composite grain}
Though the exact composition of the interstellar dust is still uncertain, graphites and silicates are the most often used for cosmic dust models 
(see for example \citet{mathis77}; \citet{draine84}). We have checked the extinction of graphite and amorphous carbon (AMC) as possible candidates 
for explaining the UV feature at 2175\AA~. Figure 3 shows the extinction curve for very small AMC and graphite grain of radius $a$=0.01$\mu m$. It is seen that the AMC does not show any peak at 2175\AA~, whereas graphite prominently shows this feature. Amorphous carbon is also highly absorbing at very long wavelengths and would provide most of the extinction longword of 0.3$\mu m$ (3.3$\mu m^{-1}$) as seen by \citet{draineIAU89} and \citet{Weig01}. Grain models with AMC are also not favored
by \citet{zubko04}. Instead, large polycyclic aromatic hydrocarbons (PAHs) molecules are likely candidates
as carriers of the 2175\AA~ feature -- a natural extension of the graphite hypothesis \citep{joblin92,li2001}. \citet{clayton03PAH} have also considered PAHs as one of the constituents in the
dust model to explain the interstellar extinction in the UV.

\begin{figure}[!ht]
\centering
\includegraphics[height=7.4cm]{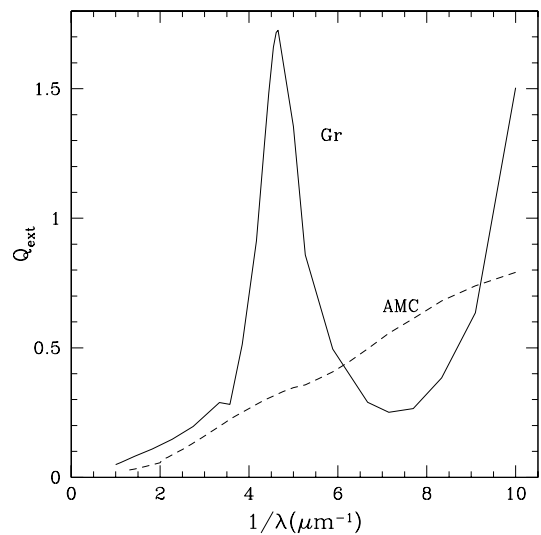}
     \caption{Extinction efficiencies for amorphous carbon (AMC) and graphite grains for very small grain size of 0.01$\mu m$ is shown in this figure. The peak in graphite curve at spectral wavelength 2175\AA~ explains why it is being used as inclusion in our composite grain model whereas no such peak is seen in AMC curve at 2175\AA~.}
\end{figure}

We calculate the extinction efficiencies $Q_{ext}$ for a composite grain model consisting of a host silicate spheroid along with graphite inclusions with the number of dipoles being $N=9640$, 14440 and 25896. The extinction efficiencies are calculated for target which are prolate spheroid in our case. The volume fractions, $f$ of the graphite inclusions in the composite grain is varied as f=0.1, 0.2 and 0.3. The extinction efficiencies for the composite grain model having number of dipoles $N=9640$ (AR= 1.33) with the variation in the volume fraction of graphite inclusion are shown in Fig. 4. The variation of extinction efficiencies for the composite grain model of  the number of dipoles $N=14440$ (AR = 2.0) and $N=25896$ (AR = 1.50) with the variation in the volume fraction of graphitic inclusions can be seen in Fig. 5 and 6 respectively.
It is clearly noted that the extinction efficiencies and the shape of the extinction
curves vary considerably as the grain size increases. The 2175\AA~ feature is clearly seen for 
small grains, viz. a=0.01$\mu m$ and 0.05$\mu m$, whereas for the larger sizes (a=0.1$\mu m$ and 0.2$\mu m$),
the feature disappears. For both the models, we see a shift in the peak 
wavelength at 2175\AA~ as the volume fraction of the inclusion increases. Further, 
the extinction efficiency is seen to vary with the variation in the volume 
fraction of graphite inclusion.

\begin{figure}[!ht]
\centering
\includegraphics[height=7.4cm]{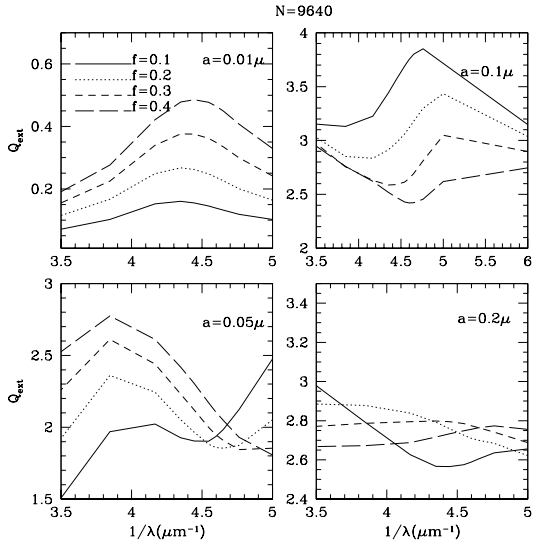}
    \caption{The figure shows extinction efficiencies for a composite grain model with number of dipoles, $N=9640$ for volume fractions of graphite inclusions, f=0.1, 0.2, 0.3 and 0.4. These extinction curves clearly show a shift in the peak wavelength `2175\AA~' (4.57$\mu m^{-1}$) and variation in extinction efficiency as the volume fraction of graphite varies. It is also to be noted that the `2175\AA~' feature vanishes for large grains with a=0.2$\mu m$}
\end{figure}

\begin{figure}[!ht]
\centering
\includegraphics[height=7.4cm]{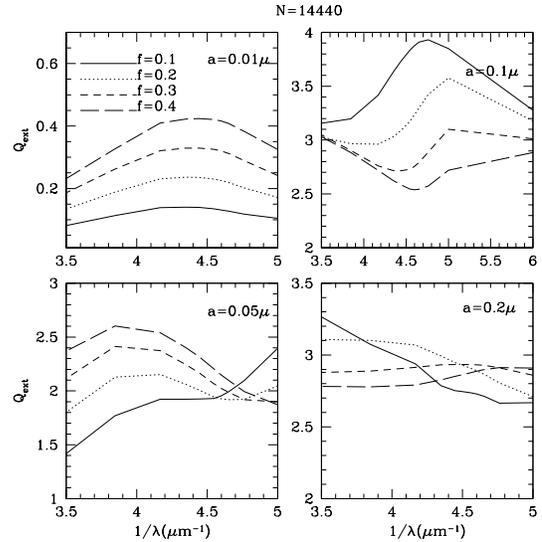}
\caption{The figure shows extinction efficiencies for a composite grain model with number of dipoles $N=14440$. The shift in the peak wavelength and variation in extinction efficiency with the volume fraction variation of graphite is seen.}
\end{figure}

\begin{figure}[!ht]
\centering
\includegraphics[height=7.4cm]{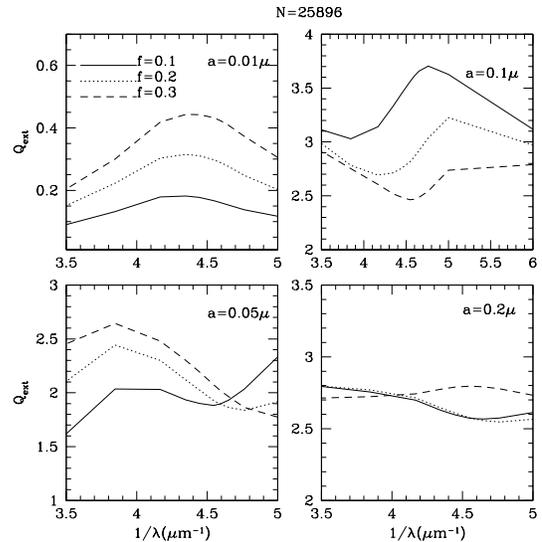}
\caption{In this figure, the extinction efficiencies for a composite grain model with number of dipoles $N=25896$ and volume fraction of graphite $f=0.1$, 0.2 and 0.3 is shown. A shift in the peak wavelength and variation in extinction efficiency as the volume fraction of graphite varies is seen.}
\end{figure}

\subsection{Interstellar extinction curve}

The interstellar extinction curve (i.e. the variation of the extinction with wavelength)
is usually expressed by the ratio: $E(\lambda-V)/E(B-V)$ vs $1/\lambda$. A power law for the grain size distribution, n(a) $\sim$ $a^{-3.5}$ \citep{mathis77}; where $a_{min}< a <a_{max}$ is used for evaluating the interstellar
extinction curve for a given grain size distribution. We calculate the extinction efficiencies of the grain models using the above power law. It must be noted that we have used two types of size distribution
(i) a=0.001-0.100$\mu m$ (denoted as $a100$ henceforth) and (ii) a=0.005-0.250 $\mu m$
(denoted as  $a250$ henceforth). 
Earlier, we have used the porous \citep{vaidya97,vaidya99} and the composite spheroidal grain models \citep{gupta07} to interpret the average
observed extinction curve in the wavelength range 0.1$\mu m$-3.4$\mu m$ \citep{gupta07}.
In this paper, we use the composite spheroidal grain models to interpret the observed
extinction in the UV, in several directions towards individual stars, selected from
various galactic environments \citep{fitz90,valencic04}.
Subsequently, in case of composite grain models, each interstellar extinction curve of the observed IUE star is compared with the model curve formed from a $\chi^2$ minimized
and best fit linear combination of the composite grains (contributory fraction p)
and solid graphite grains (contributory fraction q). By varying $p$ and $q$ each from 0.1 to 1.0 in steps of 0.1, a set of 20 model curves are generated and on comparing these model curves with the observed extinction curve of the stars, a set of reduced $\chi^2$ are obtained. A minimum $\chi^2$ from this set is chosen depending on the linear combination of $p$ and $q$. Hence, we obtain a net model interstellar extinction curve as a result of the linear combination of $p$ and $q$ which gives a minimum $\chi^2$ value. We use the following formula to obtain the set of reduced $\chi^2$\citep{beving}: 

$$
{\chi{^2_j}} = \frac {\sum_{i=1}^n (S_{i}^j-T_{i}^k)^2} {pp}
$$

where pp is the number of degrees of freedom, $S_{i}^{j}(\lambda_{i})$
is the $j$th model curve for the
corresponding $p$ and $q$ linear combination of the composite grains and bare graphite
grains and $T_{i}^{k}(\lambda_{i})$ is for the observed curve,
$\lambda_{i}$ are the wavelength points with i=1,n for n=12 wavelength points of the extinction curves. 

Tables 4 shows the best fit parameters along with the minimized $\chi^{2}$ values for the composite grain model using DDA for 26 IUE stars.

\begin{table*}
\caption{Best fit $\chi^{2}$ values and other parameters for different composite grain models generated using DDA technique.}
\small
\begin{tabular}{l c c c c c c  }
\hline
HD \#&$\chi^{2}$&p &q &N&$f_{Gr}$&a$(\mu m)$\\
\hline
 239693& 0.1552& 0.2& 0.4  &  14440& 0.1& 0.001-0.100 \\
 185418& 0.2032& 0.1& 0.6  &  14440& 0.2& 0.001-0.100 \\
 123335& 0.3719& 0.2& 0.4  &  14440& 0.1& 0.005-0.250 \\
 18352 & 0.0535& 0.2& 0.5  &  14440& 0.1& 0.005-0.250 \\
 54439 & 0.4001& 0.1& 0.4  &  9640 & 0.1& 0.001-0.100  \\
 179406& 0.2239& 0.3& 0.3  &  14440& 0.1& 0.001-0.100 \\
 24432 & 0.3149& 0.4& 0.4  &  9640 & 0.1& 0.001-0.100   \\
 217086& 0.1722& 0.2& 0.4  &  14440& 0.1& 0.001-0.100\\
 46660 & 0.1488& 0.1& 0.5  &  14440& 0.1& 0.001-0.100 \\
 281159& 0.1477& 0.2& 0.4  &  14440& 0.1& 0.001-0.100 \\
 21483 & 0.1714& 0.3& 0.3  &  9640 & 0.1& 0.001-0.100 \\
 53974 & 0.0963& 0.2& 0.3  &  9640& 0.1& 0.005-0.250 \\
 38131 & 0.2747& 0.5& 0.3  &  14440& 0.1& 0.005-0.250 \\
 217061& 0.0552& 0.2& 0.4  &  14440& 0.1& 0.005-0.250 \\
 205794& 0.1625& 0.1& 0.4  &  9640& 0.1& 0.005-0.250 \\
 46202 & 0.2304& 0.4& 0.4  &  14440& 0.1& 0.005-0.250 \\
 216658& 0.2200& 0.4& 0.4  &  14440& 0.1& 0.005-0.250 \\
 149452& 0.0778& 0.4& 0.4  &  14440& 0.1& 0.005-0.250 \\
 34078 & 0.0920& 0.3& 0.4  &  14440& 0.1& 0.005-0.250 \\
 37367 & 0.1760& 0.1& 0.5  &  14440& 0.1& 0.005-0.250 \\
 252325& 0.2172& 0.2& 0.4  &  14440& 0.3& 0.005-0.250 \\
 147701& 0.0926& 0.3& 0.2  &  9640& 0.1& 0.005-0.250 \\
 147889& 0.0652& 0.2& 0.4  &  14440& 0.3& 0.005-0.250 \\
 37903 & 0.0852& 0.3& 0.3  &  14440& 0.1& 0.005-0.250 \\
 37061 & 0.1372& 0.2& 0.2  &  14440& 0.3& 0.005-0.250 \\
 93222 & 0.0969& 0.3& 0.2  &  14440& 0.3& 0.005-0.250 \\
\hline                    
\end{tabular}             
\end{table*}

Fig. 7 and 8 shows the comparison of the observed interstellar extinction curve with the best fit model for composite grains generated using DDA technique. {From Table 1 \& 4 it is seen that the grain models with the size
distribution a=0.001-0.100 $\mu m$ fit the observed extinction curves towards stars with low $R_{v}$ (~2-3), whereas, stars with high $R_{v}$ (~4-6) fit the grains with the sizes distribution, a= 0.005-0.250$\mu m$.}

\begin{figure}[!ht]
\centering
\includegraphics[height=11.4cm]{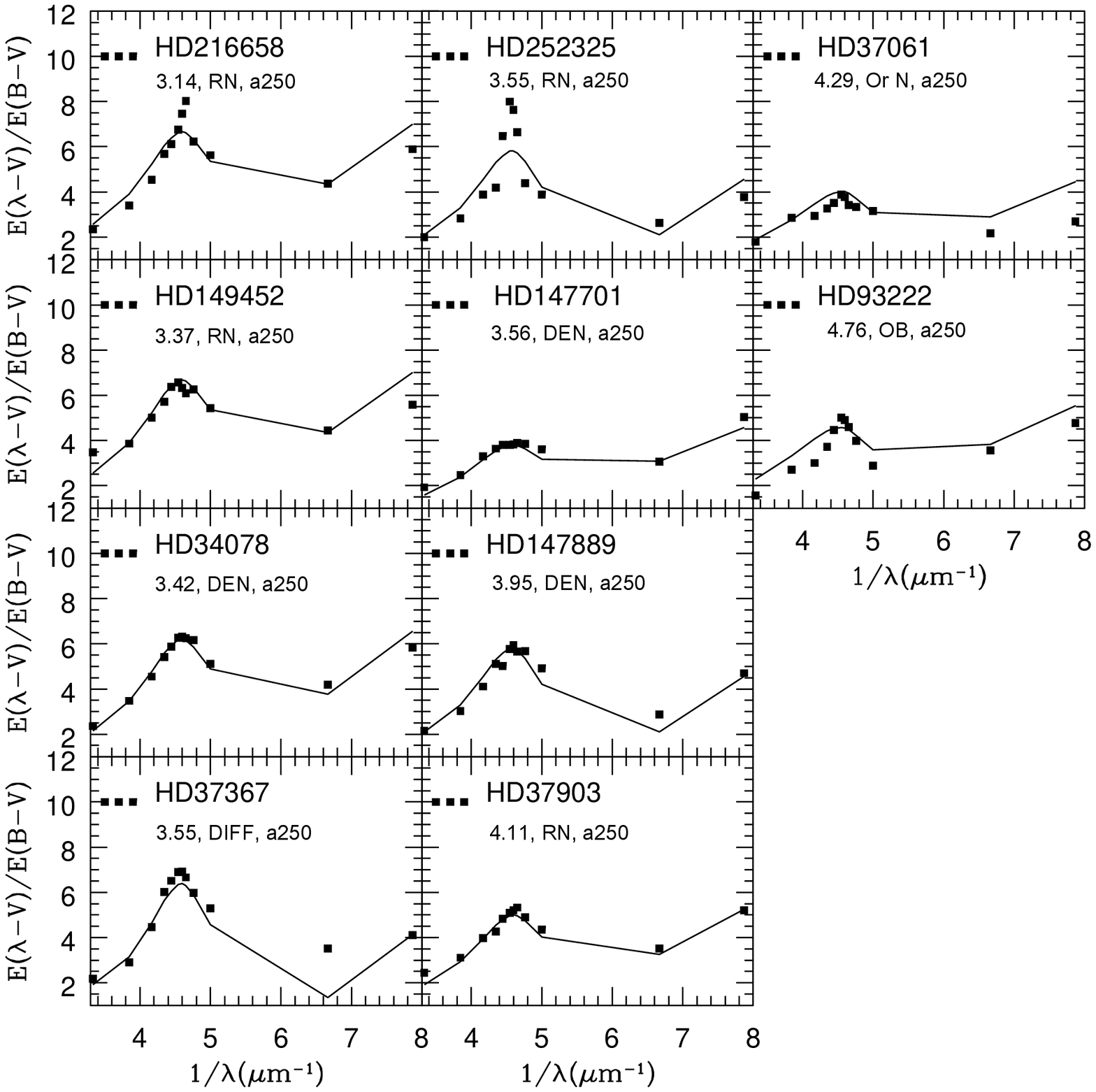}
\caption{Comparison of the observed interstellar extinction curves with the best fit composite grain model extinction curves (generated using DDA) in the wavelength range 3.17-7.87 $\mu m^{-1}$ (3200-1200\AA~). The observed $R_{v}$, environment type and the best fit grain size distribution for the sample is mentioned inside the figure.} 
\end{figure}

\begin{figure}[!ht]
\centering
\includegraphics[height=11.4cm]{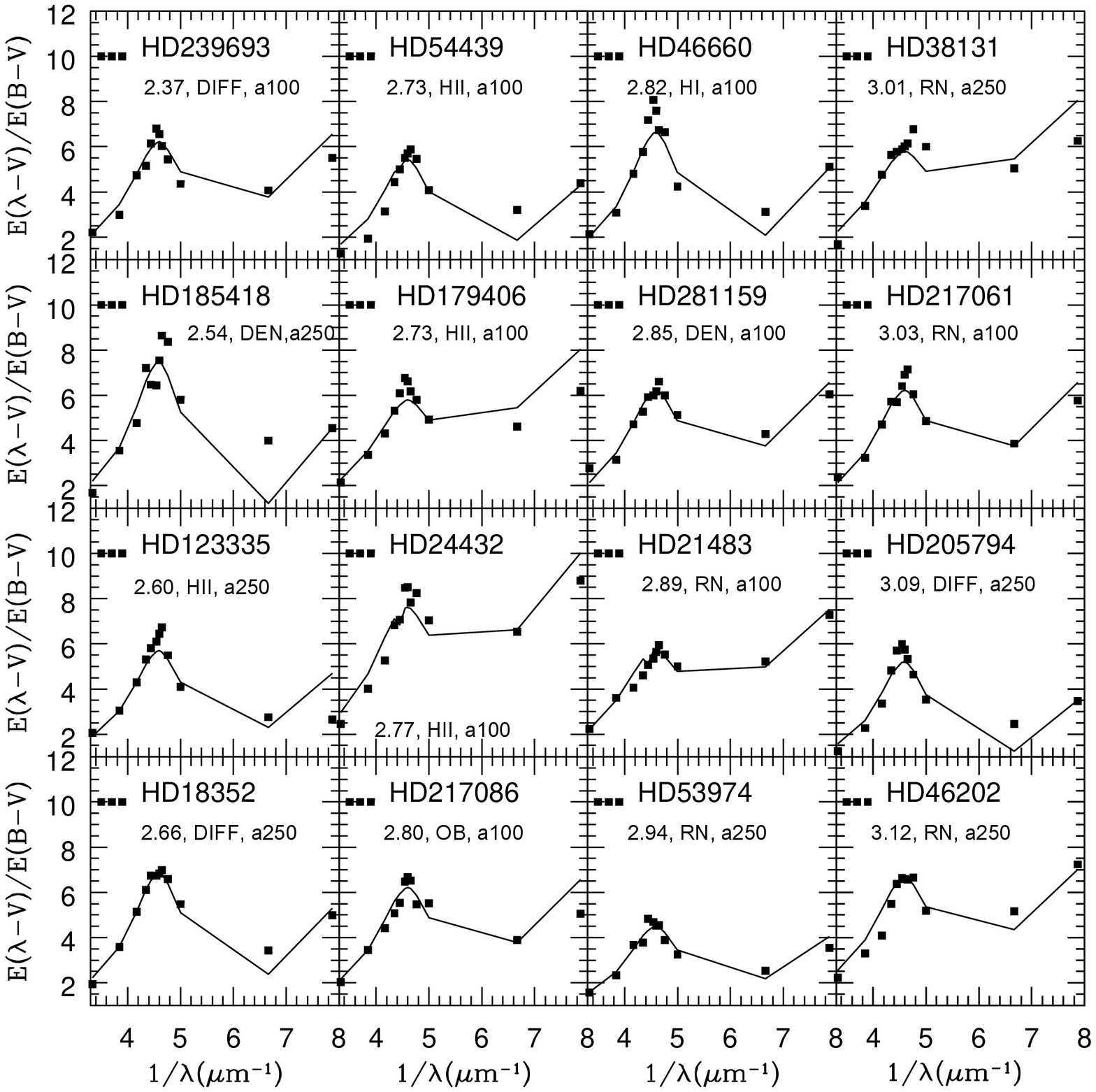}
\caption{Comparison of the observed interstellar extinction curves with the best fit 
composite grain model extinction curves (generated using DDA) in the wavelength range 3.17-7.87 $\mu m^{-1}$ (3200-1200 \AA~). The observed $R_{v}$, environment type and the best fit grain size distribution for the sample is mentioned inside the figure. }
\end{figure}

Our results on the composite spheroidal grain models i.e Table 4, Fig 7 and 8 show the best fit parameters; size distributions 0.001-0.1 $\mu m$ ($a100$) and 0.005-0.250 $\mu m$ ($a250$), shape-axial ratio; 1.33-2.0 and the composition-volume fraction of the graphite inclusions $f=0.1$, 0.2 and 0.3; for the grains in the interstellar medium towards the 26 selected stars as observed by the IUE satellite.

\subsection{Environmental effects}

In order to examine how the extinction properties are influenced by the various dust environments,
we have analyzed the extinction curves for stars in seven different galactic environments; as shown in Table 1. The variation in the strength and width of the 2175\AA~ feature is seen for various environments
i.e. from dense regions and reflection nebulae to diffuse clouds ( Fig. 7 and 8). It can be clearly seen that the dust in the dense quiescent environments and reflection nebulae produces broad bumps of larger widths 
whereas those stars lying in the diffuse environment produces narrower bumps of average widths. 
Stars around HII regions and/or which are a part of OB association produces bump of 
average widths with weak bumps. This results are in accordance with \citet{massa86}. They have shown that the observed width of the bump is strongly subject to environmental 
influences by calculating the widths of the bump and the area under the bump through the 
analytical parameterization scheme. 

In Fig. 7 and 8, we show the fitting of the extinction curves for stars in the
HII region (HII), reflection nebula (RN) and dense medium (DEN + DC) with our models. The ratio, $R_{v}$ also varies from 2.37 for HD239693 to 4.76 for HD93222. These curves further highlight star to star variation in the extinction, demonstrating  the sensitivity to local conditions. In particular, a large variation in the strength and width of the 2175\AA~ feature is seen in the extinction curves for the stars lying in the dense region i.e HD37903 ($R_{v}$=4.11), HD37061 ($R_{v}$=4.29) and HD93222 ($R_{v}$=4.76). The shape of the extinction curves for the stars in the HII region (Fig. 7 and 8) shows variation in the spectral region, shortward of 1500\AA~. In particular, see the steep rise in the extinction for HD24432 ($R_{v}$=2.77). This star lying in the diffuse clouds is fitted by the composite dust grain model of N=9640 with $a100$ size distribution. Thus, each extinction curve contains unique information about the grains along its sight line. Using analytic parameterization method, Fitzpatrick and Massa (1986, 1988) have also shown that the observed width of the 2175\AA~ feature is strongly subject to the environmental influences.

Our results on the composite grains show that the parameter $R_{v}$ varies from $\sim$2 for small grains (a=0.01$\mu m$) to $\sim$5 for the larger grains (a=0.2$\mu m$). These results also show consistency for the denser medium where $R_{v}$ has a small value (presence of small grains) and for the diffuse regions where $R_{v}$ has high value (presence of larger grains). These results are also consistent with the strength of the 2175\AA  feature viz., for the larger grains (a=0.2$\mu m$), the feature is suppressed. See for example, Fig. 7 in \citet{gupta07}.

We would like to discuss here a distinct contribution of the type of media to the spectral band feature viz., the bump at 2175\AA~. We have seen that contribution to this bump feature is due to the presence of very small graphite grains. The size of the grain shows a tremendous effect on the extinction cross sections. Hence, weakening of the bump feature can be attributed to the removal of very small grains from the dust population of the media for example; clumpy media and those of unresolved sources which are classified as extended and inhomogeneous media. Using effective medium theory (Bruggeman mixing rule), \citet{kruegel1994} have shown variation in 2175\AA~ bump feature and flattening of UV extinction curve for fluffy dust aggregates of silicate, carbon and ice with increasing grain sizes.

Extinction measured for an region is directly related to the optical depth along the line of sight. For few of the sample stars with clumpy and dense molecular media, a weakening of 2175\AA~ peak along with flattening of extinction curve is seen. This effect is attributed to the influence of scattering on the extinction properties, specially the bump feature, of the stars. \citet{natta1984} have observed a suppression in the 2175\AA~ peak followed by a flattening of far UV curve, with increasing optical depth of the media. They have also shown that inhomogeneous layer of high optical depth (high $R_{v}$) tends to produce gray extinction. 

Similar studies for low optical media have been conducted by \citet{kruegel2009}. They have investigated the influence of scattering on the extinction curve of stars. They have computed effective optical depth $\tau_{eff}$ for a variety of idealized geometrical configurations (spheres, slabs and blocks) for varying optical depth $\tau$ and analyzed the dependencies of effective optical thickness $\tau_{eff}$ on the various measurable optical properties of the dust including $\tau$. They also found out that standard dust is sensitive to spatial resolution and the structure of the medium (clumpiness, foreground/background). The extinction cross sections calculated by them, taking into account the scattering effects, for clumpy, homogeneous media and spatially unresolved stars show marked differences to the standard reddening curve.

\citet{mathis89} have fitted certain sight lines ($\rm R_{v}$=3.02, 4.83) 
using effective medium approximation (Bruggeman mixing rule).
They used composite grains (silicates and amorphous carbon) and obtained large size 
grains as the best fit parameter for sight lines with higher $\rm R_{v}$ 
values. \citet{wolff93} have used composite grains to model interstellar 
polarization towards eight lines of sight. 
They have used DDA to model the composite grains. However, their composite grain model 
with silicates and amorphous carbon/ organic refractory material failed to reproduce 
the observed polarization curve.

Several other groups have presented studies on size and composition of dust grains in interstellar medium using various techniques. For example, \citet{zubko04} have presented a dust model consisting of various components: 
PAH's, bare silicates \& AMC as well as composite particles containing silicate, 
organic refractory material, water ice and voids. They have used the method of 
regularization (MR) to solve for the optimal grain size distribution of each 
dust component knowing the observational constraints and the dust constituents 
and properties. \citet{clayton03PAH} have employed a modified version of the MEM fitting algorithm, developed by \citet{kim94} to fit the observed extinction in eight preferred sightlines/directions. They have used a 3-component homogeneous spherical grain model consisting of silicates, graphite and AMC as well as composite grain model consisting of pyroxenes, AMC and vacuum. \citet{clayton03PAH} adopted solar abundances and used EMT (extension of Bruggeman rule) to compute the optical constants of composite grains. With the 3-component homogeneous grain, they obtained the upper size cutoffs of 0.3$\mu m$  for graphite  and  AMC  and 1$\mu m$  for silicate
 grains. With the composite grain models, \citet{clayton03PAH} obtained the fit to the average observed galactic extinction curve with 0.80 (Solar) Si (Pyroxene) and 0.36 (Solar) C abundance and found the upper size cutoff size for composite grain to be 1$\mu m$. Clearly, both these grain models of \citet{clayton03PAH} show deficit of small silicate and graphite grains. On the other hand, we have found all the fits with smaller size cutoffs of 0.100$\mu m$ and 0.250$\mu m$ as compared to size cutoffs of \citet{clayton03PAH}. Although \citet{clayton03MEM} have used AMC as a third component, we did not use it since
AMC exhibits absorption at about 2500\AA~. It is also highly absorbing
at very long wavelengths and thus would provide most of the extinction longword of 0.3$\mu m$ \citep{draineIAU89}. Recently, \citet{gordan09} have analyzed 
FUSE+IUE extinction curves for 75 sightlines and have compared these curves with three different 
dust grain models given by \citet{Weig01,clayton03MEM} and \citet{zubko04}. \citet{gordan09} found that 
the models of \citet{clayton03MEM} and \citet{zubko04} provide much better fits than \citet{Weig01} model.

It is clear that the variation in the grain size distribution subject to the variation in the environment indicates that the small sized grains coagulate onto large grains in relatively dense environments, as expected \citep{draine1985,draine1990}. \citet{mathis89} have fitted certain sight lines ($\rm R_{v}$=3.02, 4.83) 
using effective medium approximation (Bruggeman mixing rule). They used composite grains (silicates and amorphous carbon) and obtained large size grains as the best fit parameter for sight lines with higher $\rm R_{v}$ 
values. \citet{Wein} have fitted a specific case of extinction toward HD21021 with small value of $R_{v}$=2.1 with a small grain size distribution of graphite/silicate grains using a simple functional fitting form. 

Shape of the grain is an important factor in determing the interstellar extinction. \citet{gupta05} have calculated the extinction efficiency for various shapes of silicate and graphitic spheroidal grains such as oblates, prolates and spheres using T-matrix theory. They have very well described the considerable variation in the extinction due to the different axial ratio grains as compared to the simple case of sphere. They also find out the best fit for explaining the observed extinction is obtained with a grain size distribution a=0.005-0.250$\mu m$ having an axial ratio of AR=1.33. However, in this work, we have used a more realistic composite dust grain model generated using DDA. We find out that most of the observed directions are well fitted by axial ratio (AR) equal to 2.0. Hence, we conclude that shape of the grain has an important role in determing the observed extinction properties of stars observed by IUE satellite. 

The composite grain models with silicate as host material and graphite inclusions, presented in this study is found to fit the observed extinction curves towards the stars lying in various interstellar environments. It must also be emphasized here that we have used more realistic DDA method to calculate the extinction efficiencies for the spheroidal composite grains. It must also be noted that \citet{perrin90}; \citet{sivan90}, \citet{wolff94,wolff98}, \citet{gupta07} and \citet{vaidya11} have shown that DDA is more accurate than the EMT based grain models.  

We plan to use the composite grain model with other carboneous materials as inclusions such as PAHs or SiC \citep{Weig01,clayton03PAH} for obtaining better fits in the UV region, 1500\AA~-1200\AA~.We also plan to interpret the extinction towards some more stars observed by the IUE satellite.

\section{CONCLUSIONS}

We have used more realistic DDA method to calculate the extinction efficiencies of the spheroidal composite grains
made up of the host silicate and graphite inclusions in the wavelength region of 1200\AA~3200\AA~. We have then, used the extinction efficiencies of the composite grains for a power law size distribution \citep{mathis77} to evaluate the interstellar extinction curves in the wavelength range 1200\AA~-3200\AA~. In the present study, we have used two size distributions viz. (i) a=0.001- 0.100 $\mu m$ and (ii) a=0.005- 0.250 $\mu m$. These extinction curves for the spheroidal composite grains are compared with the observed extinction curves obtained from the IUE
satellite data to infer the parameters such as size, shape and composition of grains. The important implications of the obtained results in terms of these physical parameters (as compared to the earlier studies) are discussed in the previous section. This study made use of a more sophisticated technique for modeling a composite dust model with various parameters that are able to characterize the actual physical dust parameters for a sample of stars, lying in different interstellar environments. The main conclusions of our study are as follows:

(i) The extinction properties of the composite grains vary considerably with 
the variation in the volume fraction of the inclusions. In particular, the extinction peak at `2175\AA~' shifts and broadens with variation in the graphite inclusions. 

(ii) The composite spheroidal grain models, with axial ratios 1.33 and 2.0 and
volume fraction of inclusions $f=0.1-0.3$, fit the observed extinction curves
reasonably well.

(iii) The ratio $\rm R_{v}=A(V)/E(B-V)$ is seen to be well correlated with the 
`2175\AA~' feature. From the sample of 26 IUE stars, those lying in the dense regions with high $R{v}$ (~4-5), show a weakening of the bump feature at 2175\AA~ followed by a flattening of far UV extinction curve whereas stars in the diffuse interstellar medium with low $R_{v}$ (~2-3) show a distinct bump at this particular wavelength. This study clearly indicates, how the extinction properties of the grains vary with the optical depth of the media (which is related to $R_{v}$) and also the grain size. It is to be noted that scattering off many unresolved stellar sources also flattens the extinction curve at this wavelength.

These results are consistent as suggested earlier by \citet{natta1984}, \citet{kruegel1994} and \citet{kruegel2009}.

In this study, we have presented the composite grain model, consisting of host silicate and graphite as inclusions and have used the results obtained for these composite grain model to infer the size distributions, shape of the grain and volume fraction of the graphite inclusions, of the interstellar dust towards 26 stars situated in the various interstellar environments. Further the composite grain models, presented in this paper, simultaneously explain the observed interstellar extinction \citep{vaidya01,gupta07}, infrared emission from the circumstellar dust \citep{vaidya11}, scattering by the cometary dust \citep{gupta06} and cosmic abundances \citep{gupta07}.

\section{Acknowledgments}

The authors acknowledge the ISRO-RESPOND project (No. ISRO/RES/2/2007-08) for funding this research.
------------------------------------------
\def\aj{AJ}%
\def\actaa{Acta Astron.}%
\def\araa{ARA\&A}%
\def\apj{ApJ}%
\def\apjl{ApJ}%
\def\apjs{ApJS}%
\def\ao{Appl.~Opt.}%
\def\apss{Ap\&SS}%
\def\aap{A\&A}%
\def\aapr{A\&A~Rev.}%
\def\aaps{A\&AS}%
\def\azh{AZh}%
\def\baas{BAAS}%
\def\bac{Bull. astr. Inst. Czechosl.}%
\def\caa{Chinese Astron. Astrophys.}%
\def\cjaa{Chinese J. Astron. Astrophys.}%
\def\icarus{Icarus}%
\def\jcap{J. Cosmology Astropart. Phys.}%
\def\jrasc{JRASC}%
\def\mnras{MNRAS}%
\def\memras{MmRAS}%
\def\na{New A}%
\def\nar{New A Rev.}%
\def\pasa{PASA}%
\def\pra{Phys.~Rev.~A}%
\def\prb{Phys.~Rev.~B}%
\def\prc{Phys.~Rev.~C}%
\def\prd{Phys.~Rev.~D}%
\def\pre{Phys.~Rev.~E}%
\def\prl{Phys.~Rev.~Lett.}%
\def\pasp{PASP}%
\def\pasj{PASJ}%
\def\qjras{QJRAS}%
\def\rmxaa{Rev. Mexicana Astron. Astrofis.}%
\def\skytel{S\&T}%
\def\solphys{Sol.~Phys.}%
\def\sovast{Soviet~Ast.}%
\def\ssr{Space~Sci.~Rev.}%
\def\zap{ZAp}%
\def\nat{Nature}%
\def\iaucirc{IAU~Circ.}%
\def\aplett{Astrophys.~Lett.}%
\def\apspr{Astrophys.~Space~Phys.~Res.}%
\def\bain{Bull.~Astron.~Inst.~Netherlands}%
\def\fcp{Fund.~Cosmic~Phys.}%
\def\gca{Geochim.~Cosmochim.~Acta}%
\def\grl{Geophys.~Res.~Lett.}%
\def\jcp{J.~Chem.~Phys.}%
\def\jgr{J.~Geophys.~Res.}%
\def\jqsrt{J.~Quant.~Spec.~Radiat.~Transf.}%
\def\memsai{Mem.~Soc.~Astron.~Italiana}%
\def\nphysa{Nucl.~Phys.~A}%
\def\physrep{Phys.~Rep.}%
\def\physscr{Phys.~Scr}%
\def\planss{Planet.~Space~Sci.}%
\def\procspie{Proc.~SPIE}%
\let\astap=\aap
\let\apjlett=\apjl
\let\apjsupp=\apjs
\let\applopt=\ao
\bibliographystyle{apj}
\bibliography{ms.bbl}

\begin{thebibliography}{70}
\expandafter\ifx\csname natexlab\endcsname\relax\def\natexlab#1{#1}\fi
\expandafter\ifx\csname href\endcsname\relax
  \def\href#1#2{}\fi
\expandafter\ifx\csname urllinklabel\endcsname\relax
  \def\urllinklabel{[LINK]}\fi
\expandafter\ifx\csname adsurllinklabel\endcsname\relax
  \def\adsurllinklabel{[ADS]}\fi

\bibitem[{{Aiello} {et~al.}(1988){Aiello}, {Barsella}, {Chlewicki},
  {Greenberg}, {Patriarchi}, \& {Perinotto}}]{aiello}
{Aiello}, S., {Barsella}, B., {Chlewicki}, G., {Greenberg}, J.~M.,
  {Patriarchi}, P., \& {Perinotto}, M. 1988, in ESA Special Publication, Vol.
  281, ESA Special Publication, 223--226


\bibitem[{{Bevington}(1969)}]{beving}
{Bevington}, P.~R. 1969, {Data reduction and error analysis for the physical
  sciences}, ed. {Bevington, P.~R.}


\bibitem[{{Brownlee}(1987)}]{brown1987}
{Brownlee}, D.~E. in , Astrophysics and Space Science Library, Vol. 134,
  Interstellar Processes, ed. D.~J. {Hollenbach}H.~A. {Thronson}, Jr., 513--530


\bibitem[{{Cardelli} \& {Clayton}(1991)}]{cardelli91}
{Cardelli}, J.~A. \& {Clayton}, G.~C. 1991, \aj, 101, 1021


\bibitem[{{Cardelli} {et~al.}(1989){Cardelli}, {Clayton}, \&
  {Mathis}}]{clayton89}
{Cardelli}, J.~A., {Clayton}, G.~C., \& {Mathis}, J.~S. 1989, \apj, 345, 245


\bibitem[{{Clayton} {et~al.}(2003{\natexlab{a}}){Clayton}, {Gordon}, {Salama},
  {Allamandola}, {Martin}, {Snow}, {Whittet}, {Witt}, \&
  {Wolff}}]{clayton03PAH}
{Clayton}, G.~C., {Gordon}, K.~D., {Salama}, F., {Allamandola}, L.~J.,
  {Martin}, P.~G., {Snow}, T.~P., {Whittet}, D.~C.~B., {Witt}, A.~N., \&
  {Wolff}, M.~J. 2003{\natexlab{a}}, \apj, 592, 947


\bibitem[{{Clayton} \& {Mathis}(1988)}]{clayton88}
{Clayton}, G.~C. \& {Mathis}, J.~S. 1988, \apj, 327, 911


\bibitem[{{Clayton} {et~al.}(2003{\natexlab{b}}){Clayton}, {Wolff}, {Sofia},
  {Gordon}, \& {Misselt}}]{clayton03MEM}
{Clayton}, G.~C., {Wolff}, M.~J., {Sofia}, U.~J., {Gordon}, K.~D., \&
  {Misselt}, K.~A. 2003{\natexlab{b}}, \apj, 588, 871


\bibitem[{Dobbie(1999)}]{dobbie99}
Dobbie, J. 1999, PhD. Thesis, Dalhousie University


\bibitem[{{Draine}(1989)}]{draineIAU89}
{Draine}, B. 1989, in IAU Symposium, Vol. 135, Interstellar Dust, ed.
  {L.~J.~Allamandola \& A.~G.~G.~M.~Tielens}, 313--+


\bibitem[{{Draine}(1985)}]{draine1985}
{Draine}, B.~T. in , Protostars and Planets II, ed. D.~C. {Black}M.~S.
  {Matthews}, 621--640


\bibitem[{{Draine}(1987)}]{draine87}
{Draine}, B.~T. 1987, \apjs, 64, 505


\bibitem[{{Draine}(1988)}]{draine88}
---. 1988, \apj, 333, 848


\bibitem[{{Draine}(1990)}]{draine1990}
{Draine}, B.~T. 1990, in Astronomical Society of the Pacific Conference Series,
  Vol.~12, The Evolution of the Interstellar Medium, ed. L.~{Blitz}, 193--205


\bibitem[{{Draine} \& {Anderson}(1985)}]{draine85}
{Draine}, B.~T. \& {Anderson}, N. 1985, \apj, 292, 494


\bibitem[{{Draine} \& {Flatau}(2003)}]{draineflat03}
{Draine}, B.~T. \& {Flatau}, P.~J. 2003, ArXiv Astrophysics e-prints


\bibitem[{{Draine} \& {Flatau}(2008)}]{manual7}
---. 2008, ArXiv e-prints


\bibitem[{{Draine} \& {Lee}(1984)}]{draine84}
{Draine}, B.~T. \& {Lee}, H.~M. 1984, \apj, 285, 89


\bibitem[{{Draine} \& {Malhotra}(1993)}]{draine93}
{Draine}, B.~T. \& {Malhotra}, S. 1993, \apj, 414, 632


\bibitem[{{Dwek}(1997)}]{dwek97}
{Dwek}, E. 1997, \apj, 484, 779


\bibitem[{{Fitzpatrick} \& {Massa}(1986)}]{massa86}
{Fitzpatrick}, E.~L. \& {Massa}, D. 1986, \apj, 307, 286


\bibitem[{{Fitzpatrick} \& {Massa}(1988)}]{massa88}
---. 1988, \apj, 328, 734


\bibitem[{{Fitzpatrick} \& {Massa}(1990)}]{fitz90}
---. 1990, \apjs, 72, 163


\bibitem[{{Fitzpatrick} \& {Massa}(2009)}]{fitz09}
---. 2009, \apj, 699, 1209


\bibitem[{{Gordon} {et~al.}(2009){Gordon}, {Cartledge}, \&
  {Clayton}}]{gordan09}
{Gordon}, K.~D., {Cartledge}, S., \& {Clayton}, G.~C. 2009, \apj, 705, 1320


\bibitem[{{Greenberg} \& {Chlewicki}(1983)}]{green83}
{Greenberg}, J.~M. \& {Chlewicki}, G. 1983, \apj, 272, 563


\bibitem[{{Gupta} {et~al.}(2005){Gupta}, {Mukai}, {Vaidya}, {Sen}, \&
  {Okada}}]{gupta05}
{Gupta}, R., {Mukai}, T., {Vaidya}, D.~B., {Sen}, A.~K., \& {Okada}, Y. 2005,
  \aap, 441, 555


\bibitem[{{Gupta} {et~al.}(2006){Gupta}, {Vaidya}, {Bobbie}, \&
  {Chylek}}]{gupta06}
{Gupta}, R., {Vaidya}, D.~B., {Bobbie}, J.~S., \& {Chylek}, P. 2006, \apss,
  301, 21


\bibitem[{{Iat{\`i}} {et~al.}(2004){Iat{\`i}}, {Giusto}, {Saija}, {Borghese},
  {Denti}, {Cecchi-Pestellini}, \& {Aiello}}]{iati04}
{Iat{\`i}}, M.~A., {Giusto}, A., {Saija}, R., {Borghese}, F., {Denti}, P.,
  {Cecchi-Pestellini}, C., \& {Aiello}, S. 2004, \apj, 615, 286


\bibitem[{{Jenniskens} \& {Greenberg}(1993)}]{green93}
{Jenniskens}, P. \& {Greenberg}, J.~M. 1993, \aap, 274, 439


\bibitem[{{Joblin} {et~al.}(1992){Joblin}, {Leger}, \& {Martin}}]{joblin92}
{Joblin}, C., {Leger}, A., \& {Martin}, P. 1992, \apjl, 393, L79


\bibitem[{{Katyal} {et~al.}(2011){Katyal}, {Gupta}, \& {Vaidya}}]{katyal11}
{Katyal}, N., {Gupta}, R., \& {Vaidya}, D.~B. 2011, Earth, Planets, and Space,
  63, 1041


\bibitem[{{Kim} {et~al.}(1994){Kim}, {Martin}, \& {Hendry}}]{kim94}
{Kim}, S., {Martin}, P.~G., \& {Hendry}, P.~D. 1994, \apj, 422, 164


\bibitem[{{Kruegel} \& {Siebenmorgen}(1994)}]{kruegel1994}
{Kruegel}, E. \& {Siebenmorgen}, R. 1994, \aap, 288, 929


\bibitem[{{Kr{\"u}gel}(2009)}]{kruegel2009}
{Kr{\"u}gel}, E. 2009, \aap, 493, 385


\bibitem[{{Li} \& {Draine}(2001)}]{li2001}
{Li}, A. \& {Draine}, B.~T. 2001, in Bulletin of the American Astronomical
  Society, Vol.~33, American Astronomical Society Meeting Abstracts, 1451


\bibitem[{{Li} \& {Greenberg}(1998)}]{greenli98}
{Li}, A. \& {Greenberg}, J.~M. 1998, \aap, 331, 291


\bibitem[{{Malloci} {et~al.}(2008){Malloci}, {Mulas}, {Cecchi-Pestellini}, \&
  {Joblin}}]{malloci2008}
{Malloci}, G., {Mulas}, G., {Cecchi-Pestellini}, C., \& {Joblin}, C. 2008,
  \aap, 489, 1183


\bibitem[{{Massa} {et~al.}(1984){Massa}, {Savage}, \& {Cassinelli}}]{massa84}
{Massa}, D., {Savage}, B.~D., \& {Cassinelli}, J.~P. 1984, \apj, 287, 814


\bibitem[{{Massa} {et~al.}(1983){Massa}, {Savage}, \& {Fitzpatrick}}]{massa83}
{Massa}, D., {Savage}, B.~D., \& {Fitzpatrick}, E.~L. 1983, \apj, 266, 662


\bibitem[{{Mathis}(1996)}]{mathis96}
{Mathis}, J.~S. 1996, \apj, 472, 643


\bibitem[{{Mathis} {et~al.}(1977){Mathis}, {Rumpl}, \& {Nordsieck}}]{mathis77}
{Mathis}, J.~S., {Rumpl}, W., \& {Nordsieck}, K.~H. 1977, \apj, 217, 425


\bibitem[{{Mathis} \& {Whiffen}(1989)}]{mathis89}
{Mathis}, J.~S. \& {Whiffen}, G. 1989, \apj, 341, 808


\bibitem[{{Meyer} \& {Savage}(1981)}]{meyer81}
{Meyer}, D.~M. \& {Savage}, B.~D. 1981, \apj, 248, 545


\bibitem[{{Natta} \& {Panagia}(1984)}]{natta1984}
{Natta}, A. \& {Panagia}, N. 1984, \apj, 287, 228


\bibitem[{{Ossenkopf}(1991)}]{Ossenkopf91}
{Ossenkopf}, V. 1991, \aap, 251, 210


\bibitem[{{Perrin} \& {Lamy}(1990)}]{perrin90}
{Perrin}, J. \& {Lamy}, P.~L. 1990, \apj, 364, 146


\bibitem[{{Perrin} \& {Sivan}(1990)}]{sivan90}
{Perrin}, J. \& {Sivan}, J. 1990, \aap, 228, 238


\bibitem[{{Purcell} \& {Pennypacker}(1973)}]{purcell73}
{Purcell}, E.~M. \& {Pennypacker}, C.~R. 1973, \apj, 186, 705


\bibitem[{{Siebenmorgen} {et~al.}(2013){Siebenmorgen}, {Voshchinnikov}, \&
  {Bagnulo}}]{sieb2013}
{Siebenmorgen}, R., {Voshchinnikov}, N.~V., \& {Bagnulo}, S. 2013, ArXiv
  e-prints


\bibitem[{{Stecher}(1965)}]{stecher65}
{Stecher}, T.~P. 1965, \apj, 142, 1683


\bibitem[{{Stecher}(1969)}]{stecher69}
---. 1969, \apjl, 157, L125+


\bibitem[{{Stecher} \& {Donn}(1965)}]{stecherdonn65}
{Stecher}, T.~P. \& {Donn}, B. 1965, \apj, 142, 1681


\bibitem[{{Vaidya} \& {Gupta}(1997)}]{vaidya97}
{Vaidya}, D.~B. \& {Gupta}, R. 1997, \aap, 328, 634


\bibitem[{{Vaidya} \& {Gupta}(1999)}]{vaidya99}
---. 1999, \aap, 348, 594


\bibitem[{{Vaidya} \& {Gupta}(2011)}]{vaidya11}
---. 2011, \aap, 528, A57+


\bibitem[{{Vaidya} {et~al.}(2001){Vaidya}, {Gupta}, {Dobbie}, \&
  {Chylek}}]{vaidya01}
{Vaidya}, D.~B., {Gupta}, R., {Dobbie}, J.~S., \& {Chylek}, P. 2001, \aap, 375,
  584


\bibitem[{{Vaidya} {et~al.}(2007){Vaidya}, {Gupta}, \& {Snow}}]{gupta07}
{Vaidya}, D.~B., {Gupta}, R., \& {Snow}, T.~P. 2007, \mnras, 379, 791


\bibitem[{{Valencic} {et~al.}(2004){Valencic}, {Clayton}, \&
  {Gordon}}]{valencic04}
{Valencic}, L.~A., {Clayton}, G.~C., \& {Gordon}, K.~D. 2004, \apj, 616, 912


\bibitem[{{Voshchinnikov} \& {Henning}(2008)}]{vosh08}
{Voshchinnikov}, N.~V. \& {Henning}, T. 2008, \aap, 483, L9


\bibitem[{{Voshchinnikov} {et~al.}(2006){Voshchinnikov}, {Il'in}, {Henning}, \&
  {Dubkova}}]{vosh06}
{Voshchinnikov}, N.~V., {Il'in}, V.~B., {Henning}, T., \& {Dubkova}, D.~N.
  2006, \aap, 445, 167


\bibitem[{{Weingartner} \& {Draine}(2001{\natexlab{a}})}]{Wein}
{Weingartner}, J.~C. \& {Draine}, B.~T. 2001{\natexlab{a}}, Apj, 548, 296


\bibitem[{{Weingartner} \& {Draine}(2001{\natexlab{b}})}]{Weig01}
---. 2001{\natexlab{b}}, \apj, 548, 296


\bibitem[{{Wolff} {et~al.}(1998){Wolff}, {Clayton}, \& {Gibson}}]{wolff98}
{Wolff}, M.~J., {Clayton}, G.~C., \& {Gibson}, S.~J. 1998, \apj, 503, 815


\bibitem[{{Wolff} {et~al.}(1994){Wolff}, {Clayton}, {Martin}, \&
  {Schulte-Ladbeck}}]{wolff94}
{Wolff}, M.~J., {Clayton}, G.~C., {Martin}, P.~G., \& {Schulte-Ladbeck}, R.~E.
  1994, \apj, 423, 412


\bibitem[{{Wolff} {et~al.}(1993){Wolff}, {Clayton}, \& {Meade}}]{wolff93}
{Wolff}, M.~J., {Clayton}, G.~C., \& {Meade}, M.~R. 1993, \apj, 403, 722


\bibitem[{{Wu} {et~al.}(1983){Wu}, {Ake}, {Boggess}, {Bohlin}, {Imhoff},
  {Holm}, {Levay}, {Panek}, {Schiffer}, \& {Turnrose}}]{wu}
{Wu}, C., {Ake}, T.~B., {Boggess}, A., {Bohlin}, R.~C., {Imhoff}, C.~L.,
  {Holm}, A.~V., {Levay}, Z.~G., {Panek}, R.~J., {Schiffer}, III, F.~H., \&
  {Turnrose}, B.~E. 1983, NASA IUE Newsl., No.~22, 2+324 pp., 22


\bibitem[{{Xiang} {et~al.}(2011){Xiang}, {Li}, \& {Zhong}}]{xiang2011}
{Xiang}, F.~Y., {Li}, A., \& {Zhong}, J.~X. 2011, \apj, 733, 91


\bibitem[{{Zubko} {et~al.}(2004{\natexlab{a}}){Zubko}, {Dwek}, \&
  {Arendt}}]{zubko94}
{Zubko}, V., {Dwek}, E., \& {Arendt}, R.~G. 2004{\natexlab{a}}, \apjs, 152, 211


\bibitem[{{Zubko} {et~al.}(2004{\natexlab{b}}){Zubko}, {Dwek}, \&
  {Arendt}}]{zubko04}
---. 2004{\natexlab{b}}, \apjs, 152, 211


\end{thebibliography}

\end{document}